\documentclass{article}

     \usepackage[preprint]{neurips_2023}

\usepackage[utf8]{inputenc} 
\usepackage[T1]{fontenc}    
\usepackage{hyperref}       
\usepackage{url}            
\usepackage{booktabs}       
\usepackage{amsfonts}       
\usepackage{nicefrac}       
\usepackage{microtype}      
\usepackage{xcolor}         
\usepackage{graphicx}
\usepackage{caption}
\usepackage{subcaption}

\title{Team IELAB at TREC Clinical Trial Track 2023: \\ Enhancing Clinical Trial Retrieval with Neural Rankers and Large Language Models}

\author{%
  Shengyao Zhuang \\
  CSIRO \\
  \texttt{shengyao.zhuang@csiro.au} \\
   \And
  Bevan Koopman \\
CSIRO \\
\texttt{bevan.koopman@csiro.au} \\
   \And
  Guido Zuccon \\
The University of Queensland \\
\texttt{g.zuccon@uq.edu.au} \\
}

\begin{document}
\maketitle

\begin{abstract}
We describe team ielab from CSIRO and The University of Queensland’s approach to the 2023 TREC Clinical Trials Track. Our approach was to use neural rankers but to utilise Large Language Models to overcome the issue of lack of training data for such rankers. Specifically, we employ ChatGPT to generate relevant patient descriptions for randomly selected clinical trials from the corpus. This synthetic dataset, combined with human-annotated training data from previous years, is used to train both dense and sparse retrievers based on PubmedBERT. Additionally, a cross-encoder re-ranker is integrated into the system. To further enhance the effectiveness of our approach, we prompting GPT-4 as a TREC annotator to provide judgments on our run files. These judgments are subsequently employed to re-rank the results. This architecture tightly integrates strong PubmedBERT-based rankers with the aid of SOTA Large Language Models, demonstrating a new approach to clinical trial retrieval.
\end{abstract}

\section{Introduction}
Participants of the TREC clinical trial track were tasked with retrieving clinical trials from \textit{ClinicalTrials.gov~\footnote{\url{https://clinicaltrials.gov/}}} given patient descriptions as queries. Our team approach this problem through a multi-stage pre-trained language model-based retriever and re-ranker pipeline. This approach has proved effective across various information retrieval tasks~\cite{nogueira2019multi,nogueira-etal-2020-document,yates-etal-2021-pretrained,gao2021rethink}, and we were interested in investigating its potential in the context of clinical trial retrieval.

Specifically, our final proposed clinical trial retrieval pipeline involves a PubmedBERT-base~\footnote{\url{https://huggingface.co/microsoft/BiomedNLP-PubMedBERT-base-uncased-abstract-fulltext}}~\citep{tinn2023fine} SPLADEv2 sparse retriever~\citep{formal2021splade} and a bi-encoder dense retriever, which then formed a hybrid first-stage retriever. The top 1000 retrieved clinical trials were then re-ranked by a PubmedBERT-large~\footnote{\url{https://huggingface.co/microsoft/BiomedNLP-PubMedBERT-large-uncased-abstract}} cross-encoder, with the scores interpolated with the first-stage hybrid retriever. Finally, we used GPT-4 to evaluate the top 20 clinical trials from our multi-stage ranking pipeline with three relevance grades. The top 20 clinical trials were then re-ranked by the GPT-4 judgments, with ties keeping the original relative orders.

During our development, we confronted two significant challenges: 
\begin{itemize}
	\item Challenge 1: Limited availability of training data. We only used patient descriptions and judgments from the previous two years of the TREC clinical trial track to train and evaluate our system. This dataset comprises 75 queries with approximately 5k gold-relevant (with relevant grade 2) description-trial pairs for training (from year 2021) and 50 queries for evaluation (from year 2022). The number of training pairs is considerably fewer than the common training data size for training BERT-based retrievers and re-rankers.
	
	\item Challenge 2: Dealing with semi-structured patient descriptions. In previous years, patient descriptions were in free natural language text. However, this year, they are in semi-structured XML data. Using previous years' data as training data may lead to a data distribution mismatch for training and inference, which could result in sub-optimal performance.
\end{itemize}

In this next section, we describe the full pipeline of our system, as well as our solutions to the above challenges.

\section{Methods}

\subsection{Training data generation}
All the retrievers and re-rankers in our system were BERT language model-based ranking models, which required training data. As mentioned in the introduction, we were only able to extract around 5k training data points from previous years' TREC clinical trials track. We believe this is insufficient for training an effective ranker. Therefore, we generated extra training data points by prompting the OpenAI \textit{gpt-3.5-turbo} model with the following one-shot example chat prompt:

\begin{quote}
	System: "You are a clinical trial specialist. You can generate patient descriptions that are best suited for particular clinical trials."
	
	User: "The clinical trial: \textit{\{gold\_trial\}} \textbackslash n Generate a patient description for this clinical trial."
	
	Assistant: "\textit{\{gold\_descrption\}}"
	
	User: "The clinical trial: \textit{\{sampled\_trial\}} \textbackslash n Generate a patient description for this clinical trial."
	
	Assistant: 
\end{quote}
where the \textit{\{gold\_trial\}} and \textit{\{gold\_descrption\}} were a pair of judged trial and its matched patient description randomly sampled from the 5k training data. The \textit{\{sampled\_trial\}} was the clinical trial randomly sampled from the corpus. With this prompt we could generate synthetic patient descriptions that potentially match the given randomly sampled clinical trials. We generated around 20k description-trial pairs.

We note that sometimes \textit{gpt-3.5-turbo} may refuse to generate the desired patient descriptions, possibly due to sensitive keywords in the clinical trial triggering OpenAI's ethical mode. This noisy data may have a negative impact on the training process. To mitigate this, we simply filtered out the generated data that contains keywords such as ``sorry", ``I cannot generate", ``AI language model", or less than 30 characters." 

To this end, we simply add these 20k generated data to the 5k human-judged data, resulting in a total of 25k training data. We consistently use this training dataset to train our rankers.

\subsection{First-stage retrievers}
The first-stage retriever in our system was a sparse-dense hybrid retriever. We utilize SPLADEv2~\citep{formal2021splade} for the learned sparse retriever and a standard bi-encoder dense retriever (DR) similar to ANCE~\citep{xiong2021approximate}. 

Both retrievers were initialized with the PubmedBERT-base checkpoint. We have observed that PubmedBERT performed significantly better than the original BERT in this task. Additionally, for the Dense Retriever, we conducted Masked Language Modeling (MLM) pre-training on the clinical trial corpus before any downstream ranking task fine-tuning. We found that this approach was beneficial for the Dense Retriever, but it does not lead to improvements for SPLADEv2.

We used a multi-stage hard negative mining training pipeline~\citep{gao-callan-2022-unsupervised,wang2022simlm,zhuang2023typos} for training SPLADEv2 and DR. In this pipeline, hard negatives for the first stage training were sampled from the top 200 retrieved clinical trials using BM25, and for the second stage training, hard negatives were mined from the top 200 clinical trials retrieved by the retriever trained in the first stage. 

For both methods, we employed a contrastive loss with three negatives per positive training data point. The patient description was truncated with a maximum length of 256 tokens, and the clinical trial was truncated with a maximum length of 512 tokens where the text was formatted with sections for title, eligibility, summary, and detailed description. We used batch size of 8 for DR and batch size of 6 for SPLADEv2. The training was performed with a single A100 GPU.

After the two stages of training for both SPLADEv2 and DR, we then created the hybrid retriever by interpolating the min-max normalized scores given by SPLADEv2 and DR with weights set to 0.5 for each methods. This was follow the study conducted by~\citet{shuai2021interpolate}.

\subsection{Cross-encoder re-ranker}

We trained a cross-encoder (CE) re-ranker to re-rank the top 1000 results retrieved by our first-stage hybrid retriever. Our CE was initialized with PubmedBERT-large and trained with the LCE loss proposed by~\cite{gao2021rethink}. We sampled hard negatives from the top 200 results retrieved by the hybrid retriever. The number of negatives per positive training example was set to 7. The batch size is set to 4. To fit within the input length limitation, we truncated patient descriptions to 182 tokens and clinical trial text to 330 tokens.

We also applied score interpolation between the re-ranker score and the hybrid retriever score with the weight set to 0.9 for the re-ranker and 0.1 for the retriever. We have found that this simple interpolation method also improved the re-ranking stage effectiveness.

\subsection{GPT4 annotator}

The final step of our system involved using OpenAI GPT-4 to serve as a TREC annotator to evaluate our re-ranker runs. This was inspired by a recent study that suggests that GPT-4 could potentially outperform human annotators for relevance assessment tasks~\cite{thomas2023large}. We than re-ranked the top 20 results from the re-ranker based on the relevance labels given by the GPT-4 model.

We used the following 3-shot example chat prompt to instruct GPT-4 in judging our runs:

\begin{quote}
	System: "You are a physician trained in medical informatics. Given a patient description and a clinical trial description, you can judge if the trial is either eligible (patient meets inclusion criteria and exclusion criteria do not apply), excluded (patient meets inclusion criteria, but is excluded on the grounds of the trial's exclusion criteria), or not relevant (patient does not meet inclusion criteria)."
	
	User: "Given the patient description: \{example\_description\} \textbackslash n and the clinical trial description: \{excluded\_trial\} \textbackslash n Judge if the clinical trial is either `eligible', `excluded' or `not relevant' for the patient. "
	
	Assistant: "excluded"
	
	User: "Given the patient description: \{example\_description\} \textbackslash n and the clinical trial description: \{eligible\_trial\} \textbackslash n Judge if the clinical trial is either `eligible', `excluded' or `not relevant' for the patient. "
	
	Assistant: "eligible"
	
	User: "Given the patient description: \{example\_description\} \textbackslash n and the clinical trial description: \{not\_relevant\_trial\} \textbackslash n Judge if the clinical trial is either `eligible', `excluded' or `not relevant' for the patient. "
	
	Assistant: "not relevant"
	
	User: "Given the patient description: \{current\_description\} \textbackslash n and the clinical trial description: \{current\_trial\} \textbackslash n Judge if the clinical trial is either `eligible', `excluded' or `not relevant' for the patient. "
	
	Assistant: 
\end{quote}
where the \{example\_description\}, \{excluded\_trial\}, \{eligible\_trial\} and \{not\_relevant\_trial\} were always fixed and they were patient description and its excluded trial (relevance label 1), eligible trial (relevance label 2) and not relevant trial (relevance label 0) receptively. These examples were sampled from the training data qrel file. The \{current\_description\} and \{current\_trial\} were the pair of patient description and clinical trial that needs to be judged. We then re-rank the top-20 results from the cross encoder in descending order of the judged relevance labels provided by GPT-4. It is important to note that there may be many ties; in such cases, we retain the original order assigned by CE.

\subsection{GPT4 format converter}

The problem remaining is that the topic format this year is very different from the previous year. In previous years, the topics were presented in free natural language text. We used this type of data for training our models. However, the topics this year consist of synthetic patient descriptions based on questionnaire templates in XML format. Thus, directly using this text data to infer our model will result in a significant mismatch in training and inference data distribution, leading to suboptimal performance of the system. Hence, we need to first convert the XML data from this year into natural language free text to mitigate the mismatch problem.
Additionally, in the actual topics of this year, none of the fields are required (i.e., they may be left blank), and there is no guaranteed format for the provided responses (i.e., each field is in natural language, not structured). This semi-structured data is difficult to convert into free natural language text with rule-based methods.

Our solution to this problem is again use LLMs to convert the semi-structured topics into free natural language text. Specifically, we use the following prompt to instruct OpenAI GPT4 model to do the converting:

\begin{quote}
	User: "Covert the following patient information in XML format into a natural language description:\textbackslash n \{xml\_topic\}"
	
	Assistant:
\end{quote}
We use the about prompt to convert all XML patient descriptions this year into natural language patient descriptions.
\section{Results}
\begin{table*}[ht]
	\centering
	\caption{
		TREC CT 2022 results.
		Overall effectiveness of the models.
		The best results are highlighted in boldface.
		Superscripts denote significant differences in paired Student's t-test with $p \le 0.05$.
	}
	\resizebox{1\textwidth}{!}{
		\begin{tabular}{c|l|c|c|c}
			\toprule
			\textbf{\#}
			& \textbf{Model}
			& \textbf{NDCG@10}
			& \textbf{P@10}
			& \textbf{Recall@1000} \\ 
			\midrule
			a &
			TREC best (frocchio\_monot5\_e) &
			0.612$^{cdefghij}$\hphantom{$^{bklm}$} &
			0.508$^{cdefghij}$\hphantom{$^{bklm}$} &
			\textbf{0.740}$^{bcdefghijklm}$\hphantom{} \\
			b &
			TREC Second (DoSSIER\_5) &
			0.556$^{cdefghi}$\hphantom{$^{ajklm}$} &
			0.456$^{cdefghij}$\hphantom{$^{aklm}$} &
			0.624$^{c}$\hphantom{$^{adefghijklm}$} \\\midrule
			c & 
			BM25  &
			0.310\hphantom{$^{abdefghijklm}$} &
			0.212\hphantom{$^{abdefghijklm}$} &
			0.366\hphantom{$^{abdefghijklm}$} \\\midrule
			d &
			DR PubmedBERT BM25 HN &
			0.404$^{h}$\hphantom{$^{abcefgijklm}$} &
			0.326$^{c}$\hphantom{$^{abefghijklm}$} &
			0.555$^{c}$\hphantom{$^{abefghijklm}$} \\
			e &
			DR PubmedBERT+MLM BM25 HN  &
			0.407$^{h}$\hphantom{$^{abcdfgijklm}$} &
			0.328$^{c}$\hphantom{$^{abdfghijklm}$} &
			0.599$^{cd}$\hphantom{$^{abfghijklm}$} \\
			f &
			DR PubmedBERT+MLM DR HN &
			0.427$^{ch}$\hphantom{$^{abdegijklm}$} &
			0.324$^{c}$\hphantom{$^{abdeghijklm}$} &
			0.599$^{c}$\hphantom{$^{abdeghijklm}$} \\\midrule
			g &
			SPLADEv2 PubmedBERT BM25 HN &
			0.373\hphantom{$^{abcdefhijklm}$} &
			0.302$^{c}$\hphantom{$^{abdefhijklm}$} &
			0.576$^{c}$\hphantom{$^{abdefhijklm}$} \\
			h &
			SPLADEv2 PubmedBERT+MLM BM25 HN &
			0.351\hphantom{$^{abcdefgijklm}$} &
			0.292\hphantom{$^{abcdefgijklm}$} &
			0.558$^{c}$\hphantom{$^{abdefgijklm}$} \\
			i &
			SPLADEv2 PubmedBERT DR HN &
			0.423$^{ch}$\hphantom{$^{abdefgjklm}$} &
			0.328$^{c}$\hphantom{$^{abdefghjklm}$} &
			0.597$^{c}$\hphantom{$^{abdefghjklm}$} \\\midrule
			j &
			Hybrid f and i &
			0.502$^{cdefghi}$\hphantom{$^{abklm}$} &
			0.380$^{cfghi}$\hphantom{$^{abdeklm}$} &
			0.653$^{cdefghi}$\hphantom{$^{abklm}$} \\\midrule
			k &
			CE re-rank j &
			0.609$^{cdefghij}$\hphantom{$^{ablm}$} &
			0.480$^{cdefghij}$\hphantom{$^{ablm}$} &
			0.653$^{cdefghi}$\hphantom{$^{abjlm}$} \\
			l &
			Hybrid j and k &
			0.621$^{bcdefghij}$\hphantom{$^{akm}$} &
			0.488$^{cdefghij}$\hphantom{$^{abkm}$} &
			0.653$^{cdefghi}$\hphantom{$^{abjkm}$} \\\midrule
			m &
			GPT-4 judge l top-20 &
			\textbf{0.659}$^{bcdefghijkl}$\hphantom{$^{a}$} &
			\textbf{0.568}$^{bcdefghijkl}$\hphantom{$^{a}$} &
			0.653$^{cdefghi}$\hphantom{$^{abjkl}$} \\
			\bottomrule
		\end{tabular}
	}
	\label{tab:results}
\end{table*}

We use TREC CT 2022 queries and judgments to serve as the validation dataset and TREC CT 2023 as the test dataset to evaluate our system. In Table~\ref{tab:results}, we present our results obtained on TREC CT 2022. Our results demonstrate that DR and SPLADEv2 are significantly better than BM25. Two-stage hard negative mining training significantly improves both DR and SPLADEv2 (lines f and i). Further MLM pre-training on the target corpus only helps for DR (when comparing lines d and e), especially for Recall@1000, but does not help SPLADEv2 (when comparing lines g and h). A hybrid of the best DR and SPLADEv2 gives considerable improvement (line j).

\begin{table*}[t]
	\centering
	\caption{
		TREC CT 2023 results.
		Overall effectiveness of the models.
		The best results are highlighted in boldface.
	}
	\resizebox{0.6\textwidth}{!}{
		\begin{tabular}{c|l|c|c|c}
			\toprule
			\textbf{\#}
			& \textbf{Model}
			& \textbf{NDCG@10}
			& \textbf{P@10}
			& \textbf{Recall@1000} \\ 
			\midrule
			a & DR & 0.5768 & 0.3243 & 0.3670 \\
			b & SPLADEv2 & 0.5971 & 0.3243 & 0.3482 \\
			c & Hybrid & 0.5763 & 0.2946 & \textbf{0.3878} \\
			d & CE\_weighted & 0.6716 & 0.4432 & \textbf{0.3878} \\
			e & GPT4 & \textbf{0.7363} & \textbf{0.5108} & \textbf{0.3878} \\\bottomrule
		\end{tabular}
	}
	\label{tab:results2}
\end{table*}

\begin{figure*}[t]
	
	\begin{subfigure}{.5\textwidth}
		\centering
		\includegraphics[width=\linewidth]{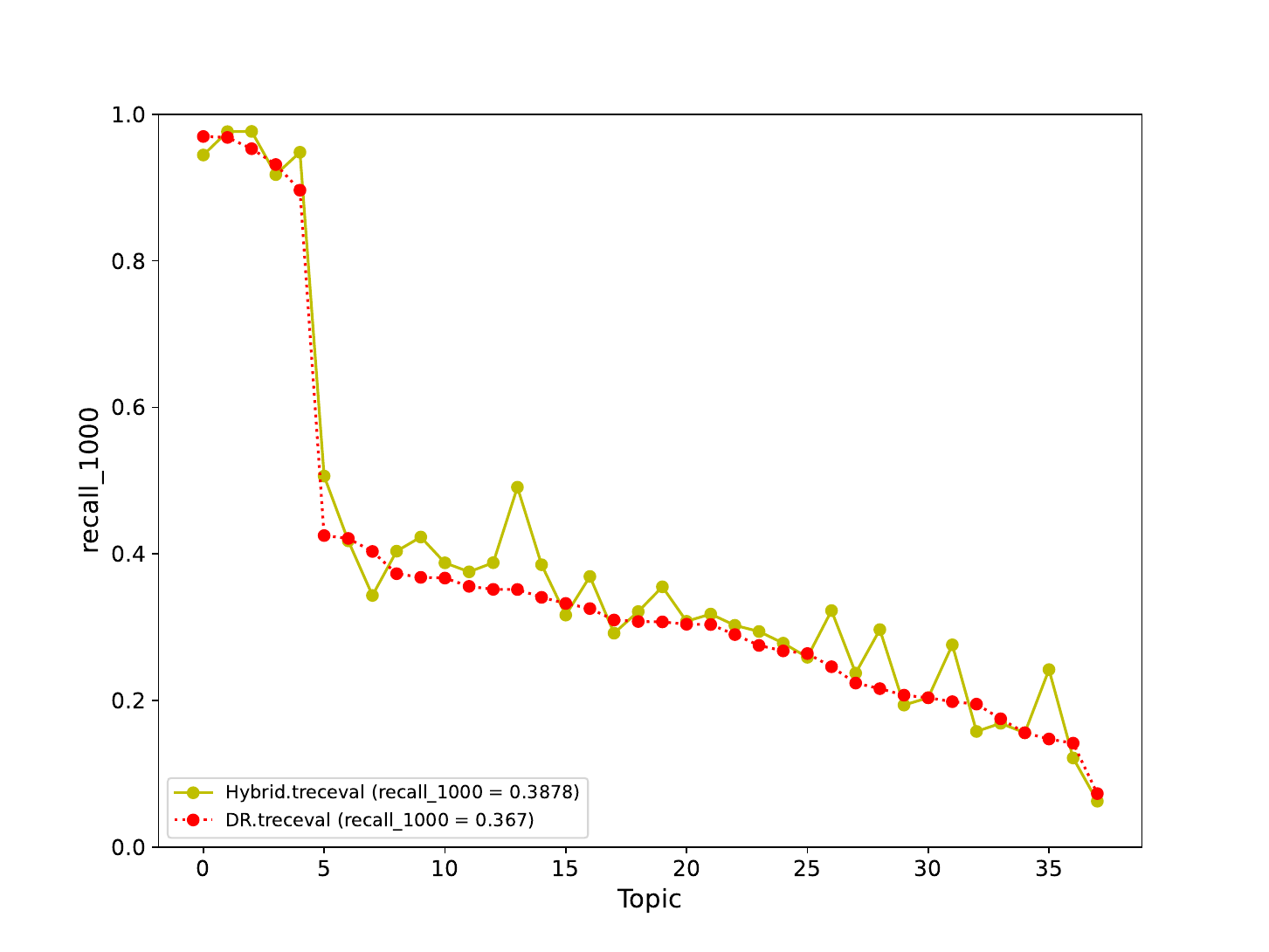}
		\caption{Hybrid retriever improvements over DR.}
		\label{fig:sfig1}
	\end{subfigure}
	\vspace{15pt}
	\begin{subfigure}{.5\textwidth}
		\centering
		\includegraphics[width=\linewidth]{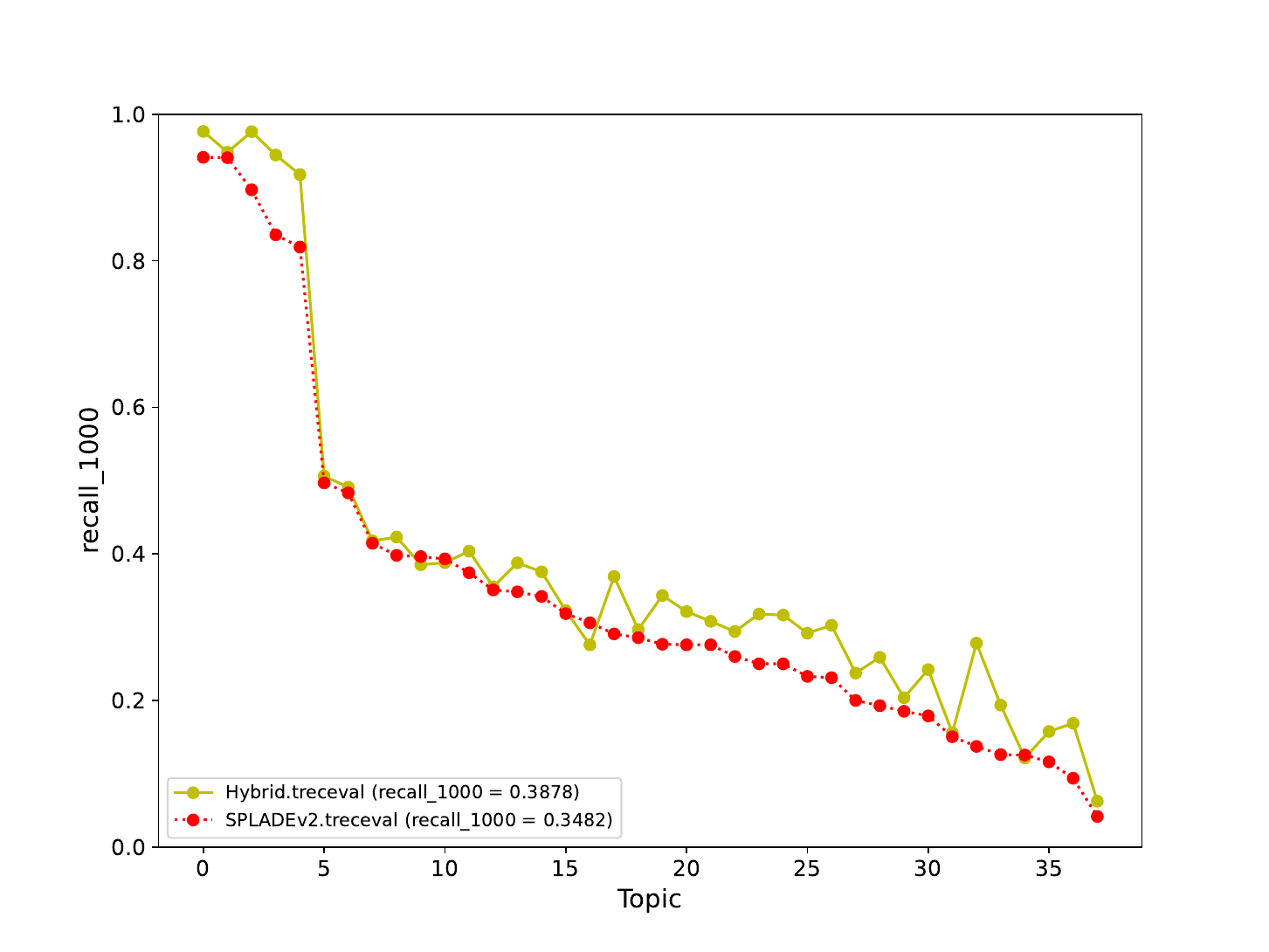}
		\caption{Hybrid retriever improvements over SPLADEv2.}
		\label{fig:sfig2}
	\end{subfigure}
	\vspace{10pt}
	\begin{subfigure}{.5\textwidth}
		\includegraphics[width=\linewidth]{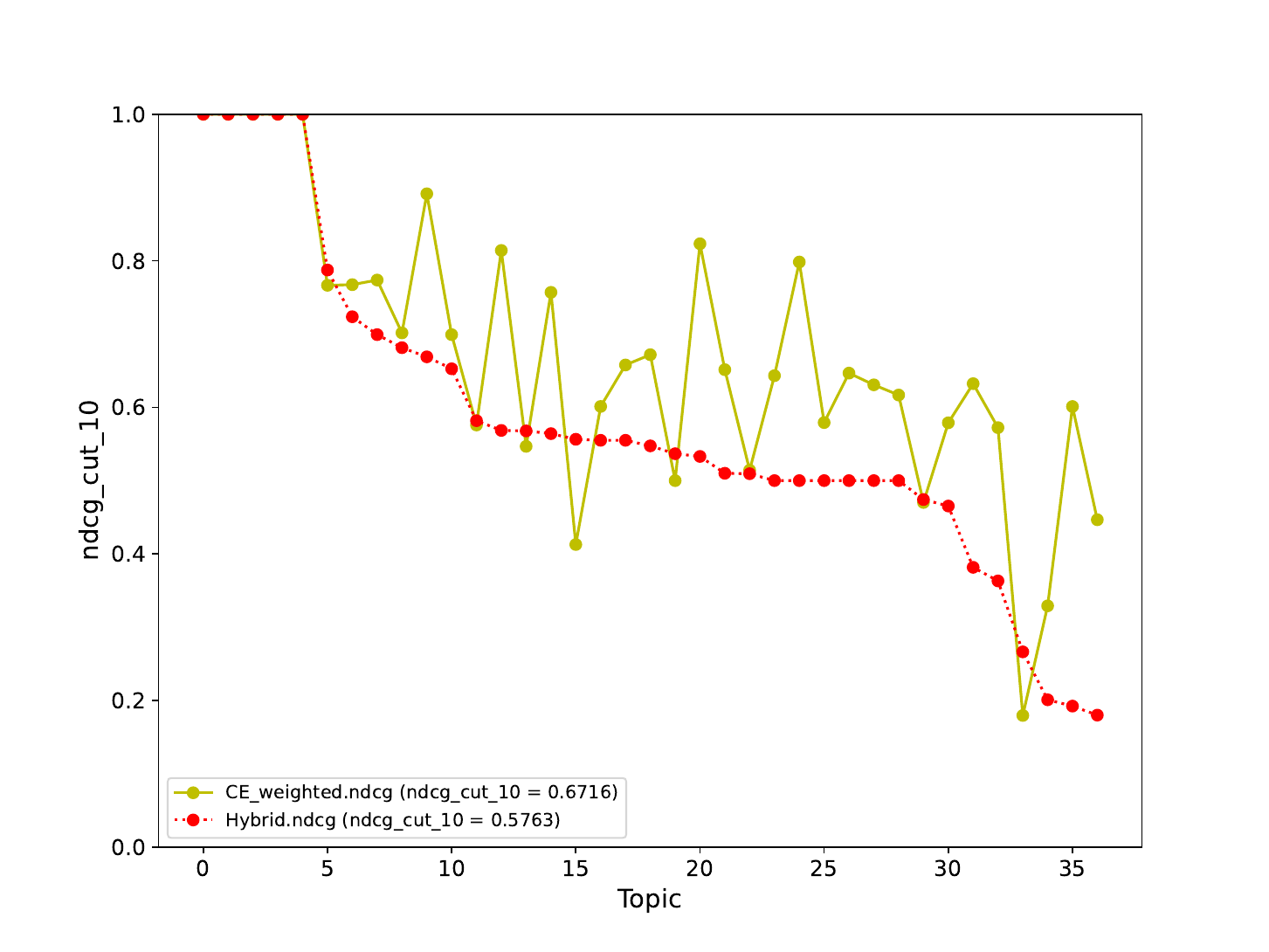}
		\caption{CE improvements over Hybrid retriever.}
		\label{fig:sfig3}
	\end{subfigure}
	\begin{subfigure}{.5\textwidth}
		\includegraphics[width=\linewidth]{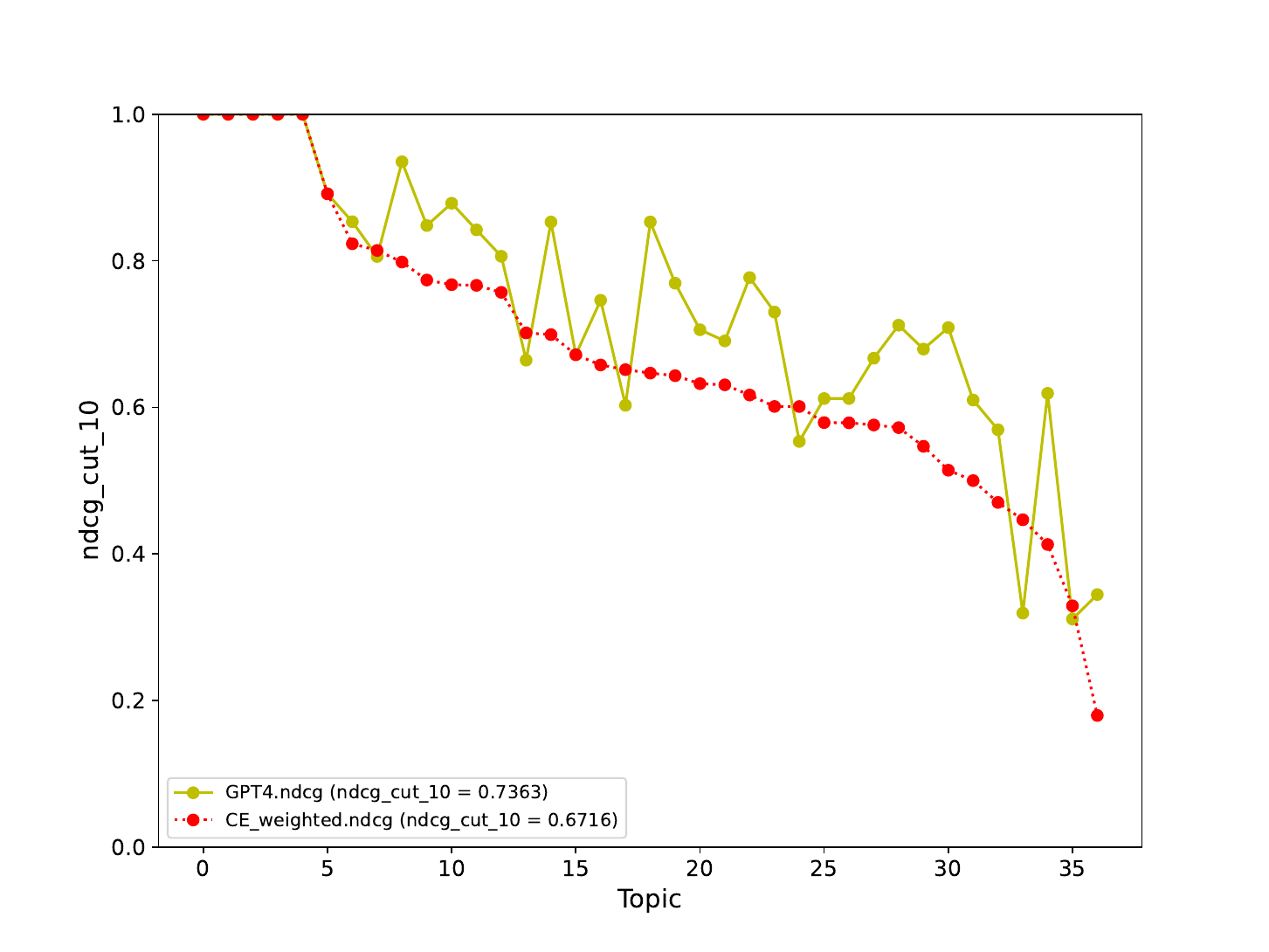}
		\caption{GPT4 improvements over CE.}
		\label{fig:sfig4}
	\end{subfigure}
	\caption{Query-by-query improvements over different stages in the pipeline.}
	\label{fig:1}
\end{figure*}

The cross-encoder re-ranker can also significantly improve by re-ranking the top 1000 results from the hybrid first stage retriever (line k). Notably, simply applying another fusion of CE results with the hybrid retriever results can result in a better NDCG@10 score than the best TREC run of that year (line l).

Finally, applying GPT-4 to judge the top 20 documents from the CE hybrid retriever system and re-ranking accordingly gives the best NDCG@10 and P@10 scores (line m). We note that there are still a lot of unjudged documents in the top 10 results of our best run, whereas the TREC runs (lines a and b) are all judged. Thus, our best run is underestimated and could possibly achieve even higher scores.

In Table~\ref{tab:results2}, we report the results we obtained for TREC CT 2023. We submitted a total of 5 runs which can represent different components of our proposed clinical trial retrieval system. They are labeled as follows: (a) DR, corresponding to line f in Table~\ref{tab:results}. (b) SPLADEv2, corresponding to line i in Table~\ref{tab:results}. (c) Hybrid, corresponding to line j in Table~\ref{tab:results}. (d) CE\_weighted, corresponding to line l in Table~\ref{tab:results}. (e) GPT4, corresponding to line m in Table~\ref{tab:results}. All the runs were judged by TREC annotators.
Our results for TREC CT 2023 generally follow the trend observed in TREC CT 2022, with the only exception being that the hybrid model yields lower NDCG@10 and P@10 scores. However, considering it is a first-stage retrieval and its top 1000 documents will be re-ranked by the CE re-ranker, achieving the best Recall@1000 is promising.

In Figure~\ref{fig:1}, we also plot query-by-query improvements of different components in our systems. As the figures illustrate, overall, the effectiveness of most of the queries can be enhanced by the component in the next step in the pipeline, which is desirable.

\section{Conclusion}
Our team introduced an innovative clinical trial retrieval system that integrates PLM-based ranking models and LLMs. Our approach minimizes the initial reliance on human-labeled data, leveraging LLMs to generate synthetic training data for the development of robust PLM-based dense and sparse retrieval models, as well as cross-encoder re-rankers. Additionally, we harness the few-shot power of LLMs in judging retrieved results, further enhancing the system's ranking effectiveness. The outcome of our participation in the TREC clinical trial track underscores the remarkable competitiveness of our multistage clinical trial retrieval pipeline.

%

\bibliography{anthology,custom}

\begin{thebibliography}{12}
\expandafter\ifx\csname natexlab\endcsname\relax\def\natexlab#1{#1}\fi

\bibitem[{Formal et~al.(2021)Formal, Lassance, Piwowarski, and
  Clinchant}]{formal2021splade}
Thibault Formal, Carlos Lassance, Benjamin Piwowarski, and St{\'e}phane
  Clinchant. 2021.
\newblock Splade v2: Sparse lexical and expansion model for information
  retrieval.
\newblock \emph{arXiv preprint arXiv:2109.10086}.

\bibitem[{Gao and Callan(2022)}]{gao-callan-2022-unsupervised}
Luyu Gao and Jamie Callan. 2022.
\newblock \href {https://doi.org/10.18653/v1/2022.acl-long.203} {Unsupervised
  corpus aware language model pre-training for dense passage retrieval}.
\newblock In \emph{Proceedings of the 60th Annual Meeting of the Association
  for Computational Linguistics (Volume 1: Long Papers)}, pages 2843--2853,
  Dublin, Ireland. Association for Computational Linguistics.

\bibitem[{Gao et~al.(2021)Gao, Dai, and Callan}]{gao2021rethink}
Luyu Gao, Zhuyun Dai, and Jamie Callan. 2021.
\newblock Rethink training of bert rerankers in multi-stage retrieval pipeline.
\newblock In \emph{European Conference on Information Retrieval}, pages
  280--286.

\bibitem[{Nogueira et~al.(2020)Nogueira, Jiang, Pradeep, and
  Lin}]{nogueira-etal-2020-document}
Rodrigo Nogueira, Zhiying Jiang, Ronak Pradeep, and Jimmy Lin. 2020.
\newblock \href {https://doi.org/10.18653/v1/2020.findings-emnlp.63} {Document
  ranking with a pretrained sequence-to-sequence model}.
\newblock In \emph{Findings of the Association for Computational Linguistics:
  EMNLP 2020}, pages 708--718, Online. Association for Computational
  Linguistics.

\bibitem[{Nogueira et~al.(2019)Nogueira, Yang, Cho, and
  Lin}]{nogueira2019multi}
Rodrigo Nogueira, Wei Yang, Kyunghyun Cho, and Jimmy Lin. 2019.
\newblock Multi-stage document ranking with bert.
\newblock \emph{arXiv preprint arXiv:1910.14424}.

\bibitem[{Thomas et~al.(2023)Thomas, Spielman, Craswell, and
  Mitra}]{thomas2023large}
Paul Thomas, Seth Spielman, Nick Craswell, and Bhaskar Mitra. 2023.
\newblock Large language models can accurately predict searcher preferences.
\newblock \emph{arXiv preprint arXiv:2309.10621}.

\bibitem[{Tinn et~al.(2023)Tinn, Cheng, Gu, Usuyama, Liu, Naumann, Gao, and
  Poon}]{tinn2023fine}
Robert Tinn, Hao Cheng, Yu~Gu, Naoto Usuyama, Xiaodong Liu, Tristan Naumann,
  Jianfeng Gao, and Hoifung Poon. 2023.
\newblock Fine-tuning large neural language models for biomedical natural
  language processing.
\newblock \emph{Patterns}, 4(4).

\bibitem[{Wang et~al.(2022)Wang, Yang, Huang, Jiao, Yang, Jiang, Majumder, and
  Wei}]{wang2022simlm}
Liang Wang, Nan Yang, Xiaolong Huang, Binxing Jiao, Linjun Yang, Daxin Jiang,
  Rangan Majumder, and Furu Wei. 2022.
\newblock Simlm: Pre-training with representation bottleneck for dense passage
  retrieval.
\newblock \emph{arXiv preprint arXiv:2207.02578}.

\bibitem[{Wang et~al.(2021)Wang, Zhuang, and Zuccon}]{shuai2021interpolate}
Shuai Wang, Shengyao Zhuang, and Guido Zuccon. 2021.
\newblock \href {https://doi.org/10.1145/3471158.3472233} {Bert-based dense
  retrievers require interpolation with bm25 for effective passage retrieval}.
\newblock In \emph{Proceedings of the 2021 ACM SIGIR International Conference
  on Theory of Information Retrieval}, ICTIR '21, page 317–324, New York, NY,
  USA. Association for Computing Machinery.

\bibitem[{Xiong et~al.(2021)Xiong, Xiong, Li, Tang, Liu, Bennett, Ahmed, and
  Overwijk}]{xiong2021approximate}
Lee Xiong, Chenyan Xiong, Ye~Li, Kwok-Fung Tang, Jialin Liu, Paul~N. Bennett,
  Junaid Ahmed, and Arnold Overwijk. 2021.
\newblock \href {https://openreview.net/forum?id=zeFrfgyZln} {Approximate
  nearest neighbor negative contrastive learning for dense text retrieval}.
\newblock In \emph{International Conference on Learning Representations}.

\bibitem[{Yates et~al.(2021)Yates, Nogueira, and
  Lin}]{yates-etal-2021-pretrained}
Andrew Yates, Rodrigo Nogueira, and Jimmy Lin. 2021.
\newblock \href {https://doi.org/10.18653/v1/2021.naacl-tutorials.1}
  {Pretrained transformers for text ranking: {BERT} and beyond}.
\newblock In \emph{Proceedings of the 2021 Conference of the North American
  Chapter of the Association for Computational Linguistics: Human Language
  Technologies: Tutorials}, pages 1--4, Online. Association for Computational
  Linguistics.

\bibitem[{Zhuang et~al.(2023)Zhuang, Shou, Pei, Gong, Ren, Zuccon, and
  Jiang}]{zhuang2023typos}
Shengyao Zhuang, Linjun Shou, Jian Pei, Ming Gong, Houxing Ren, Guido Zuccon,
  and Daxin Jiang. 2023.
\newblock Typos-aware bottlenecked pre-training for robust dense retrieval.
\newblock \emph{arXiv preprint arXiv:2304.08138}.

\end{thebibliography}
\bibliographystyle{acl_natbib}

\end{document}